# Three-dimensional equations of generalized dynamics of 18-fold symmetry soft-matter quasicrystals

Zhi-Yi Tang1 and Tian-You Fan 2*

1 School of Computer Science and Technology, Beijing Institute of Technology, Beijing 100081, China
2 School of Physics, Beijing Institute of Technology, Beijing 100081,China
*Corresponding author, e-mail:tyfan2013@163.com

**Abstract** This letter presents a three-dimensional form of governing equations ofgeneralized dynamics of 18-fold symmetry soft-matter quasicrystals, according to the dynamics basis there are first and second phason elementary excitations apart from phonons and fluid phonon. In the derivation, the group representation theory is a key point. The complete form of the theory includes an equation of state. The governing equations present some important meaning in the study on thermodynamics of the matter which is also introduced.

## 1. Introduction

The Ref [1] reported the two-dimensional form of governing equations ofgeneralized dynamics of 18-fold symmetry soft-matter quasicrystals; monograph [2] introduced some solutions of the equations. The three-dimensional equations of generalized dynamics of the first kind of soft-matter quasicrystals were obtained by Fan and Tang [3]. One of the basis of the theory is the six-dimensional embedding space concept of Hu et al [4], this concept and group representation theory yield the elastic constitutive law of the 18-fold symmetry structure, but in two-dimensional case.Hu et al expected their theory can be used to describe the possible solid quasicrystals, although this type quasicrystals in solids have not been observed so far. It is exciting the 18-fold symmetry soft-matter quasicrystals in colloids were discovered by Fischer et al[5] after 17 years of the prediction of Ref [4]. This shows the power of group and group representation theory! After the observation, Fan[2] and his group carry out a series of work to study the quasicrystals and suggest the concept of the first and second kinds of soft-matter quasicrystals, in which the 18- and possible 7-, 9- and 14-fold symmetry soft-matter quasicrystals belong to the second kind ones. For this kind of soft-matter quasicrystals the study is in the regime of two-dimension since the publication of Ref [4].

The need of applications in basic study and applications require doing the three-dimensional analysis, in this letter we report the work, in which the theory of group representation is used, and the results are immediately used in the analysis of stability of the 18-fold symmetry soft-matter quasicrystals [6], this problem remains a question of great debate to this day due to the quite different formation mechanism between soft-matter and solid quasicrystals.

## 2. Some basic relations

According to the hypothesis of Hu et al [4] on the six- dimensional embedding space for the 7-, 9-, 14- and 18-fold symmetry quasicrsystals, there are elementary excitations phonons, first and second phasons, whose fields are u, v and w respectively. For soft-matter quasicrystals, Fan [1,2] suggested it is needed to introduce another elementary excitations---fluid phonon, the corresponding field is the fluid velocity V. To describe the deformation and motion of the matter, the tensors of phonon strain, first phason strain, second phason strain and fluid phonon deformation rate respectively as follows

$$\varepsilon_{ij} = \frac{1}{2}\left(\frac{\partial u_i}{\partial x_j} + \frac{\partial u_j}{\partial x_i}\right), v_{ij} = \frac{\partial v_i}{\partial x_j}, w_{ij} = \frac{\partial w_i}{\partial x_j}, \dot{\xi}_{ij} = \frac{1}{2}\left(\frac{\partial V_i}{\partial x_j} + \frac{\partial V_j}{\partial x_i}\right) \quad (1)$$

in which $x = x_1, y = x_2, z = x_3$, $i, j = 1, 2, 3$, and the corresponding constitutive law [1,2]

$$\left.\begin{aligned}
\sigma_{ij} &= C_{ijkl}\varepsilon_{kl} + r_{ijkl}v_{kl} + R_{ijkl}w_{kl} \\
\tau_{ij} &= T_{ijkl}v_{kl} + r_{klij}\varepsilon_{kl} + G_{ijkl}w_{kl} \\
H_{ij} &= K_{ijkl}w_{kl} + R_{klij}\varepsilon_{kl} + G_{klij}v_{kl} \\
p_{ij} &= -p\delta_{ij} + \sigma'_{ij} = -p\delta_{ij} + \eta_{ijkl}\dot{\xi}_{kl}
\end{aligned}\right\} \quad i, j, k, l = 1, 2, 3 \quad (2)$$

in which $\sigma_{ij}$ denotes the phonon stress tensor associated to the phonon strain tensor $\varepsilon_{ij}$, $C_{ijkl}$ the phonon elastic constants, $\tau_{ij}$ the first phason stress tensor associated to the first phason strain tensor $v_{ij}$, $T_{ijkl}$ the first phason elastic constants, $H_{ij}$ the second phason stress tensor associated to the second phason strain tensor $w_{ij}$, $K_{ijkl}$ the second phason elastic constants, $r_{ijkl}, r_{klij}$ the phonon-first phason coupling ($u - v$ coupling) elastic constants, $R_{ijkl}, R_{klij}$ the phonon-second phason coupling ($u - w$ coupling) elastic constants, $G_{ijkl}, G_{klij}$ the first-second phason coupling ($v - w$ coupling) elastic constants, $p_{ij}$ the fluid stress tensor, $p$ the fluid pressure, $\delta_{ij}$ the unit tensor, $\sigma'_{ij}$ the fluid viscous stress tensor, $\eta_{ijkl}$ the fluid viscous coefficient constants, respectively.

By using the theory of group representation developed by Hu et al [7] we find all of following independent nonzero elastic constants for point group 18mm of 18-fold symmetry soft-matter quasicrystals in three-dimensional case (those in two-dimensional case were found by Hu et al [7]):

$$\left.\begin{aligned}
&C_{1111} = C_{11}, C_{1122} = C_{12}, C_{3333} = C_{33}, \\
&C_{1133} = C_{13}, C_{2323} = C_{44}, C_{1212} = C_{66}, \\
&(C_{11} - C_{12})/2 = C_{66}, \\
&T_{1111} = T_{2222} = T_{2121} = T_{1212} = T_1, \\
&T_{1122} = T_{2211} = -T_{2112} = -T_{1221} = T_2, \\
&T_{2323} = T_{1313} = T_3 \\
&K_{1111} = K_{2222} = K_{2121} = K_{1212} = K_1, \\
&K_{1122} = K_{2211} = -K_{2112} = -K_{1221} = K_2, \\
&K_{2323} = K_{1313} = K_3 \\
&G_{1111} = -G_{1122} = G_{2211} = -G_{2222} = -G_{1212} = -G_{1221} = G_{2112} = G_{2121} = G
\end{aligned}\right\} \quad (3)$$

So that we have the concrete form of the constitute laws as for the phonons

$$\left.\begin{aligned}
\sigma_{xx} &= C_{11}\varepsilon_{xx} + C_{12}\varepsilon_{yy} + C_{13}\varepsilon_{zz} \\
\sigma_{yy} &= C_{12}\varepsilon_{xx} + C_{11}\varepsilon_{yy} + C_{13}\varepsilon_{zz} \\
\sigma_{zz} &= C_{13}\varepsilon_{xx} + C_{13}\varepsilon_{yy} + C_{33}\varepsilon_{zz} \\
\sigma_{yz} &= \sigma_{zy} = 2C_{44}\varepsilon_{yz} \\
\sigma_{zx} &= \sigma_{xz} = 2C_{44}\varepsilon_{zx} \\
\sigma_{xy} &= \sigma_{yx} = 2C_{66}\varepsilon_{xy}
\end{aligned}\right\} \quad (4)$$

for the first phasons

$$\left.\begin{aligned}
\tau_{xx} &= T_1 v_{xx} + T_2 v_{yy} + G(w_{xx} - w_{yy}) \\
\tau_{yy} &= T_2 v_{xx} + T_1 v_{yy} + G(w_{xx} - w_{yy}) \\
\tau_{xy} &= T_1 v_{xy} - T_2 v_{yx} - G(w_{xy} + w_{yx}) \\
\tau_{yx} &= T_1 v_{yx} - T_2 v_{xy} + G(w_{xy} + w_{yx}) \\
\tau_{xz} &= T_3 v_{xz}, \qquad \tau_{yz} = T_3 v_{yz}
\end{aligned}\right\} \quad (5)$$

for the second phasons

$$\left.\begin{aligned}
H_{xx} &= K_1 w_{xx} + K_2 w_{yy} + G(v_{xx} + v_{yy}) \\
H_{yy} &= K_2 w_{xx} + K_1 w_{yy} - G(v_{xx} + v_{yy}) \\
H_{xy} &= K_1 w_{xy} - K_2 w_{yx} - G(v_{xy} - v_{yx}) \\
H_{yx} &= K_1 w_{yx} - K_2 w_{xy} - G(v_{xy} - v_{yx}) \\
H_{xz} &= K_3 w_{xz}, \qquad H_{yz} = K_3 w_{yz}
\end{aligned}\right\} \quad (6)$$

In addition there is the constitutive law for the fluid phonon

$$\left.\begin{aligned}
p_{xx} &= -p + 2\eta\dot{\xi}_{xx} - \frac{2}{3}\eta\dot{\xi}_{kk} \\
p_{yy} &= -p + 2\eta\dot{\xi}_{yy} - \frac{2}{3}\eta\dot{\xi}_{kk} \\
p_{zz} &= -p + 2\eta\dot{\xi}_{zz} - \frac{2}{3}\eta\dot{\xi}_{kk} \\
p_{yz} &= p_{zy} = 2\eta\dot{\xi}_{yz} \\
p_{zx} &= p_{xz} = 2\eta\dot{\xi}_{zx} \\
p_{xy} &= p_{yx} = 2\eta\dot{\xi}_{xy} \\
\dot{\xi}_{kk} &= \dot{\xi}_{xx} + \dot{\xi}_{yy} + \dot{\xi}_{zz}
\end{aligned}\right\} \quad (7)$$

The detail on the treatment with theory of the group representation here is omitted due to the limitation of the space.

## 3. Three-dimensional equations of generalized dynamics of point

## group $18mm$ soft-matter quasicrystals

In Refs [1,2] a generalized dynamics of soft-matter quasicrystals is developed, i.e., based on the results in the Sec 2, and define the Hamiltonian

$$\begin{aligned}
H &= H[\Psi(\mathbf{r},t)] \\
&= \int \frac{g^2}{2\rho} d^d\mathbf{r} + \int \left[\frac{1}{2} A \left(\frac{\delta\rho}{\rho_0}\right)^2 + B \left(\frac{\delta\rho}{\rho_0}\right) \nabla\cdot\mathbf{u}\right] d^d\mathbf{r} + F_{el} \\
&= H_{kin} + H_{density} + F_{el} \\
F_{el} &= F_u + F_v + F_w + F_{uv} + F_{uw} + F_{vw}, \quad \mathbf{g} = \rho \mathbf{V}
\end{aligned} \quad (8)$$

where $\mathbf{V}$ represents the fluid velocity field mentioned above, $A, B$ the constants describing mass density variation. The last term of (8) represents elastic energies, which consists of phonons, phasons and phonon-phason coupling and phason-phason coupling parts, respectively:

$$\begin{aligned}
F_u &= \int \frac{1}{2} C_{ijkl} \varepsilon_{ij} \varepsilon_{kl} d^d\mathbf{r} \\
F_v &= \int \frac{1}{2} T_{ijkl} v_{ij} v_{kl} d^d\mathbf{r} \\
F_w &= \int \frac{1}{2} K_{ijkl} w_{ij} w_{kl} d^d\mathbf{r} \\
F_{uv} &= \int \left(r_{ijkl} \varepsilon_{ij} v_{kl} + r_{klij} v_{ij} \varepsilon_{kl}\right) d^d\mathbf{r} \\
F_{uw} &= \int \left(R_{ijkl} \varepsilon_{ij} w_{kl} + R_{klij} w_{ij} \varepsilon_{kl}\right) d^d\mathbf{r} \\
F_{vw} &= \int \left(G_{ijkl} v_{ij} w_{kl} + G_{klij} w_{ij} v_{kl}\right) d^d\mathbf{r}
\end{aligned} \quad (9)$$

There are the fundamental laws of the mass conservation

$$\frac{\partial \rho}{\partial t} + \nabla_k (\rho V_k) = 0 \quad (10)$$

the momentum conservation or the generalized Navier-Stokes equations

$$\begin{aligned}
\frac{\partial g_i(\mathbf{r},t)}{\partial t} &= -\nabla_k(\mathbf{r})(V_k g_i) + \nabla_j(\mathbf{r})\left(-p\delta_{ij} + \eta_{ijkl}\nabla_k(\mathbf{r})V_l\right) - \left(\delta_{ij} - \nabla_i u_j\right)\frac{\delta H}{\delta u_j(\mathbf{r},t)} \\
&+ (\nabla_i v_j)\frac{\delta H}{\delta v_j(\mathbf{r},t)} + (\nabla_i w_j)\frac{\delta H}{\delta w_j(\mathbf{r},t)} - \rho\nabla_i(\mathbf{r})\frac{\delta H}{\delta\rho(\mathbf{r},t)}, \quad g_j = \rho V_j
\end{aligned} \quad (11)$$

the symmetry breaking rule on the motion of phonons

$$\frac{\partial u_i(\mathbf{r},t)}{\partial t} = -V_j \nabla_j(\mathbf{r}) u_i - \Gamma_u \frac{\delta H}{\delta u_i(\mathbf{r},t)} + V_i \quad (12)$$

in which $\Gamma_u$ represents phonon dissipation coefficient, and the symmetry breaking rules on the motion of the first phasons

$$\frac{\partial v_i(\mathbf{r},t)}{\partial t} = -V_j \nabla_j(\mathbf{r}) v_i - \Gamma_v \frac{\delta H}{\delta v_i(\mathbf{r},t)} \quad (13)$$

and the second phasons

$$\frac{\partial w_i(\mathbf{r},t)}{\partial t} = -V_j \nabla_j(\mathbf{r}) w_i - \Gamma_w \frac{\delta H}{\delta w_i(\mathbf{r},t)} \quad (14)$$

in which $\Gamma_v$ and $\Gamma_w$ represents the first and second phason dissipation coefficients, respectively. However the equation system up to now is not closed yet, because the number of field variables is greater than that of field equations. We must supplement an equation, the equation of state, i.e., the relation between fluid pressure and mass density:

$$p = f(\rho)$$

which is a difficult topic in the study of soft matter.

After some probes [1] the equation of state can be taken as follows

$$p = f(\rho) = 3 \frac{k_B T}{l^3 \rho_0^3} \left( \rho_0^2 \rho + \rho_0 \rho^2 + \rho^3 \right) \quad (15)$$

where $\rho_0$ is the initial value of the mass density, or the rest mass density, $k_B$ the Boltzmann constant, $T$ the absolute temperature, $l$ the characteristic size of soft-matter quasicrystals, in general this is a meso-characteristic size, $l = 1 \sim 100\,\text{nm}$, and our computation shows if take $l = 8 \sim 9\,\text{nm}$ the computational results are in the best accuracy. In the above the superscript $d$ means the dimension of the integral domain.

Substituting (9) into (8) then into (11)-(14) and combining (10) and (15) but by omitting the higher terms $(\nabla_i u_j) \frac{\delta H}{\delta u_j(\mathbf{r},t)}, (\nabla_i v_j) \frac{\delta H}{\delta v_j(\mathbf{r},t)}$ and $(\nabla_i w_j) \frac{\delta H}{\delta w_j(\mathbf{r},t)}$ of in (11) we obtain the final governing equations of three-dimensional dynamics of 18-fold symmetry soft-matter quasicrystals as following

$$\frac{\partial \rho}{\partial t} + \nabla \bullet (\rho \mathbf{V}) = 0 \quad (16a)$$

$$\begin{aligned}
\frac{\partial (\rho V_x)}{\partial t} &+ \frac{\partial (V_x \rho V_x)}{\partial x} + \frac{\partial (V_y \rho V_x)}{\partial y} + \frac{\partial (V_z \rho V_x)}{\partial z} = -\frac{\partial p}{\partial x} + \eta \nabla^2 V_x + \frac{1}{3} \eta \frac{\partial}{\partial x} \nabla \bullet \mathbf{V} \\
&+ \left( C_{11} \frac{\partial^2}{\partial x^2} + C_{66} \frac{\partial^2}{\partial y^2} + C_{44} \frac{\partial^2}{\partial z^2} \right) u_x + (C_{12} + C_{66}) \frac{\partial^2 u_y}{\partial x \partial y} + (C_{13} + C_{44}) \frac{\partial^2 u_z}{\partial x \partial z} \quad (16b) \\
&- B \frac{\partial}{\partial x} \nabla \bullet \mathbf{u} - (A - B) \frac{1}{\rho_0} \frac{\partial \delta \rho}{\partial x}
\end{aligned}$$

$$\begin{aligned}
\frac{\partial (\rho V_y)}{\partial t} &+ \frac{\partial (V_x \rho V_y)}{\partial x} + \frac{\partial (V_y \rho V_y)}{\partial y} + \frac{\partial (V_z \rho V_y)}{\partial z} = -\frac{\partial p}{\partial y} + \eta \nabla^2 V_y + \frac{1}{3} \eta \frac{\partial}{\partial y} \nabla \bullet \mathbf{V} \\
&+ (C_{12} + C_{66}) \frac{\partial^2 u_x}{\partial x \partial y} + \left( C_{66} \frac{\partial^2}{\partial x^2} + C_{11} \frac{\partial^2}{\partial y^2} + C_{44} \frac{\partial^2}{\partial z^2} \right) u_y + (C_{13} + C_{44}) \frac{\partial^2 u_z}{\partial y \partial z} \quad (16c) \\
&- B \frac{\partial}{\partial y} \nabla \bullet \mathbf{u} - (A - B) \frac{1}{\rho_0} \frac{\partial \delta \rho}{\partial y}
\end{aligned}$$

$$\frac{\partial(\rho V_z)}{\partial t}+\frac{\partial(V_x\rho V_z)}{\partial x}+\frac{\partial(V_y\rho V_z)}{\partial y}+\frac{\partial(V_z\rho V_z)}{\partial z}=-\frac{\partial p}{\partial z}+\eta\nabla^2 V_z+\frac{1}{3}\eta\frac{\partial}{\partial z}\nabla\cdot\mathbf{V}$$
$$+(C_{13}+C_{44})\left(\frac{\partial^2 u_x}{\partial x\partial z}+\frac{\partial^2 u_y}{\partial y\partial z}\right)+\left(C_{44}\frac{\partial^2}{\partial x^2}+C_{44}\frac{\partial^2}{\partial y^2}+C_{33}\frac{\partial^2}{\partial z^2}\right)u_z \quad (16d)$$
$$-B\frac{\partial}{\partial z}\nabla\cdot\mathbf{u}-(A-B)\frac{1}{\rho_0}\frac{\partial\delta\rho}{\partial z}$$

$$\frac{\partial u_x}{\partial t}+V_x\frac{\partial u_x}{\partial x}+V_y\frac{\partial u_x}{\partial y}+V_z\frac{\partial u_x}{\partial z}=V_x$$
$$+\Gamma_{\mathbf{u}}\left[\left(C_{11}\frac{\partial^2}{\partial x^2}+C_{66}\frac{\partial^2}{\partial y^2}+C_{44}\frac{\partial^2}{\partial z^2}\right)u_x+(C_{12}+C_{66})\frac{\partial^2 u_y}{\partial x\partial y}+(C_{13}+C_{44})\frac{\partial^2 u_z}{\partial x\partial z}\right] \quad (16e)$$

$$\frac{\partial u_y}{\partial t}+V_x\frac{\partial u_y}{\partial x}+V_y\frac{\partial u_y}{\partial y}+V_z\frac{\partial u_y}{\partial z}=V_y$$
$$+\Gamma_{\mathbf{u}}\left[(C_{12}+C_{66})\frac{\partial^2 u_x}{\partial x\partial y}+\left(C_{66}\frac{\partial^2}{\partial x^2}+C_{11}\frac{\partial^2}{\partial y^2}+C_{44}\frac{\partial^2}{\partial z^2}\right)u_y+(C_{13}+C_{44})\frac{\partial^2 u_z}{\partial y\partial z}\right] \quad (16f)$$

$$\frac{\partial u_z}{\partial t}+V_x\frac{\partial u_z}{\partial x}+V_y\frac{\partial u_z}{\partial y}+V_z\frac{\partial u_z}{\partial z}=V_z$$
$$+\Gamma_{\mathbf{u}}\left[(C_{13}+C_{44})\left(\frac{\partial^2 u_x}{\partial x\partial z}+\frac{\partial^2 u_y}{\partial y\partial z}\right)+\left(C_{44}\frac{\partial^2}{\partial x^2}+C_{44}\frac{\partial^2}{\partial y^2}+C_{33}\frac{\partial^2}{\partial z^2}\right)u_z\right] \quad (16g)$$

$$\frac{\partial v_x}{\partial t}+V_x\frac{\partial v_x}{\partial x}+V_y\frac{\partial v_x}{\partial y}+V_z\frac{\partial v_x}{\partial z}=$$
$$\Gamma_v\left[\left(T_1\frac{\partial^2}{\partial x^2}+T_1\frac{\partial^2}{\partial y^2}+T_3\frac{\partial^2}{\partial z^2}\right)v_x+G\left(\frac{\partial^2}{\partial x^2}-\frac{\partial^2}{\partial y^2}\right)w_x-2G\frac{\partial^2 w_y}{\partial x\partial y}\right] \quad (16h)$$

$$\frac{\partial v_y}{\partial t}+V_x\frac{\partial v_y}{\partial x}+V_y\frac{\partial v_y}{\partial y}+V_z\frac{\partial v_y}{\partial z}=$$
$$\Gamma_v\left[\left(T_1\frac{\partial^2}{\partial x^2}+T_1\frac{\partial^2}{\partial y^2}+T_3\frac{\partial^2}{\partial z^2}\right)v_y+2G\frac{\partial^2 w_x}{\partial x\partial y}+G\left(\frac{\partial^2}{\partial x^2}-\frac{\partial^2}{\partial y^2}\right)w_y\right] \quad (16i)$$

$$\frac{\partial w_x}{\partial t}+V_x\frac{\partial w_x}{\partial x}+V_y\frac{\partial w_x}{\partial y}+V_z\frac{\partial w_x}{\partial z}=$$
$$\Gamma_{\mathbf{w}}\left[\left(K_1\frac{\partial^2}{\partial x^2}+K_1\frac{\partial^2}{\partial y^2}+K_3\frac{\partial^2}{\partial z^2}\right)w_x+G\left(\frac{\partial^2}{\partial x^2}-\frac{\partial^2}{\partial y^2}\right)v_x+2G\frac{\partial^2 v_y}{\partial x\partial y}\right] \quad (16j)$$

$$\frac{\partial w_y}{\partial t}+V_x\frac{\partial w_y}{\partial x}+V_y\frac{\partial w_y}{\partial y}+V_z\frac{\partial w_y}{\partial z}=$$
$$\Gamma_{\mathbf{w}}\left[\left(K_1\frac{\partial^2}{\partial x^2}+K_1\frac{\partial^2}{\partial y^2}+K_3\frac{\partial^2}{\partial z^2}\right)w_y-2G\frac{\partial^2 v_x}{\partial x\partial y}+G\left(\frac{\partial^2}{\partial x^2}-\frac{\partial^2}{\partial y^2}\right)v_y\right] \quad (16k)$$

$$p=f(\rho)=3\frac{k_B T}{l^3 \rho_0^3}\left(\rho_0^2\rho+\rho_0\rho^2+\rho^3\right) \quad (16l)$$

in which there are 12 equations including (16a) the mass conservation equation, (16b-16d) the momentum conservation equations or generalized Navier-Stokes equations, (16e-16g) the equations of motion of phonons due to the symmetry breaking, (16h),(16i) the dissipation equations of the first phasons, (16j),(16k) the dissipation equations of the second phasons, (16*l*) the equation of state, respectively, and the field variables are also 12: $\rho, p, V_x, V_y, V_z, u_x, u_y, u_z, v_x, v_y, w_x, w_y$, so the amount of field equations and field variables are equal to each other, the governing equations are consistent and solvable mathematically. In the equation set (16) if there is lack of the equation of state, then the set is not closed, which cannot be solved so that is no meaning physically and mathematically. This shows the importance of the equation of state.

## 4. Application example---the stability of 18-fold symmetry soft-matter quasicrystals

Because the formation mechanism of soft-matter quasicrystals is quite different from that of solid quasicrystals, the stability of 12-fold symmetry soft-matter quasicrystals is an important problem, which has been studied by Lifshitz et al [8] soon after the discovery of the matter.They and other researchers continue working up to now [9-13], and these work carried out based on so-called effective free energy suggested by Lifshitz and Petrich [14]. Fan and Tang study the problem based on an extended free energy [15,6], which could carry out not only for 12-fold symmetry soft-matter quasicrystals but also for 18-fold symmetry case, and the latter might be studied for the first time. Although our study takes quite different point of view from Lifshitz and other researchers, we can obtain the quantitative results which are very simple and intuitive,and all of quantities contained in the stability criterion can be measured by experiments. Especially the results are independent from any simulations. The above results can be obtained duo to the combination between thermodynamics and the generalized dynamics of soft-matter quasicrystals, in particular the Hamiltonian and the constitutive law of the soft-matter quasicrystals play important role. The study for stability of soft-matter quasicrystals [15,6] shows the importance of the three-dimensional theory of generalized dynamics of soft-matter quasicrystals.

## 5. Conclusion and discussion

By taking the group representation theory developed by Hu et al [7] , the principle of generalized dynamics of soft-matter quasicrystals suggested by Fan [1,2] and the six-dimensional embedding space hypothesis of Hu et al [4], we obtain the three-dimensional dynamic equations of 18-fold symmetry soft-matter quasicrystals. The successful application of the three-dimensional dynamic results to the stability of the matter shows the importance of the theory. Of course there are other applications of the theory apart from the stability problem, such as numerical solutions of some dynamical samples of 18-fold symmetry soft-matter quasicrystals, which will be informed in other reports of ours.

**Acknowledgement**The work is supported by the National Natural Science Foundation of China through grant 11272053.Zhi-Yi Tang is also grateful to the support in part of the National Natural Science Foundation of China through the grant 11871098.